\documentclass[prb,twocolumn,showpacs,superscriptaddress]{revtex4}
 \usepackage{graphicx}
 \usepackage{dcolumn}

\begin{document}

 \title{Effects of deposition dynamics on epitaxial growth}

 \author{Jikeun Seo}
 \affiliation{Department of Ophthalmic Optics, Chodang University,
 Muan 534-701, Korea }
 \author{Hye-Young Kim}
 \affiliation{ Department of Physics, The Pennsylvania State University,
 University  Park, PA 16802}
 \author{J.-S. Kim}
 \affiliation{Department of Physics, Sook-Myung Women's University,
 Seoul, 140-742,  Republic of Korea}

 \date{\today}

 \begin{abstract}
   The dynamic effects, such as the steering and the screening effects during deposition,
 on an epitaxial growth (Cu/Cu(001)), is studied by kinetic Monte
 Carlo simulation that incorporates molecular
 dynamic simulation to rigorously take the interaction of the
 deposited atom with the substrate atoms into account.
 We find three characteristic features of the surface morphology
 developed by grazing angle deposition:
 (1) enhanced surface roughness, (2) asymmetric mound, and (3)
 asymmetric slopes of mound sides.
  Regarding their dependence on both deposition angle and substrate
 temperature, a reasonable agreement of the simulated results
 with the previous experimental ones is found.
 The characteristic growth features by grazing angle deposition
 are mainly caused by the inhomogeneous deposition flux due to
 the steering and screening effects, where the steering effects play
 the major role rather than the screening effects.
 Newly observed in the present simulation is that the
 side of mound  in each direction is composed of various facets
 instead of all being in one selected mound angle even if the slope
 selection is attained, and that the slope selection does not
 necessarily mean the facet selection.
 \end{abstract}

  \pacs{PACS numbers:68.35.-p,68.37.-d }
  \maketitle

  \section{Introduction}

   For the tailored growth of a structure on a substrate, careful consideration of
 energetic parameters, such as surface and interface energies, and
 kinetic parameters, such as diffusion barriers, is necessary.
 However, the parameters related to the dynamics of the deposited
 atoms, other than the deposition flux, have not been seriously
 considered. Recently, Dijken, Jorritsma, and Poelsema\cite{Dijken1}
 has observed by spot profile analysis low energy electron
 diffraction (SPA-LEED)  that the islands formed on Cu (001) show
 rectangular symmetry in contrast to the square symmetry of the
 substrate when the 0.5 monolayer (ML) of Cu is deposited at
 deposition angle ($\theta$) of 80$^o$ from the surface normal. They
 suggest a model in which the interaction between deposited atom and
 substrate atoms steers the deposited atom and thus results in the
 inhomogeneous deposition flux, so called the steering effect. Our
 previous study on the thin film growth \cite{Seo1}, simulating
 dynamics of the deposited atoms by molecular dynamics (MD), has
 confirmed this hypothetical model. These studies demonstrate that
 the dynamic parameters involved in the deposition process, which
 have been ignored in most of the previous studies, exert notable
 effects on the thin film growth. Since the steering effect is
 unavoidable during deposition process, it influences all the thin
 film growth by atomic deposition in some degree. Even in the case of
 the normal deposition, the steering effect is found to affect the
 thin film growth.\cite{Montalenti,Montalenti2,Yu} It has been also
 shown that the deposition dynamics is a cause of the unstable growth
 of thin film on a vicinal surface.\cite{Seo2} \par

 Dijken, Jorritsma and Poelsema \cite{Dijken2} has observed, in the
 thicker film (40 ML) growth with $\theta =80^o$ that the surface
  is rougher  than that with $\theta = 0^o$. It has
 also been observed that the slope of the mound facing the deposition
 direction is much steeper than that of the side shadowed from the
 deposition.\cite{Dijken2} Although a qualitative model is suggested
 relating the experimental result with the inhomogeneous deposition
 flux due to the steering and screening effects,\cite{Wormeester}
 there has not been any realistic growth study or simulation work
 which confirms such speculation.

 In the present work, we perform kinetic Monte Carlo (KMC) simulation
 in conjunction with MD simulation that is designed to probe the
 effects of the dynamic processes on the thick film growth in atomic
 level. The main results of our simulation are as follows; (1) the
 roughness increases with increasing deposition angle, (2) the mound
 formed in the thick film growth has rectangular symmetry with sides
 elongated in the direction perpendicular to the deposition
 direction, and (3) the slopes of the illuminated and shadowed sides
 of the mound significantly differ, which is consistent with the experimental
 results.\cite{Dijken2} The aforementioned three characteristic
 morphological features are mainly caused by the inhomogeneous
 deposition flux on the top terrace of the mound mainly due to the steering
 effects rather than the screening effects. In the present study, in
 addition, it is found that the side of mound in each direction is
 formed of various local facets instead of all being in one selected
 mound angle, even if the slope selection is attained,  and that the
 experimentally observed mound slope actually corresponds to the mean
 slope of various local facets coexisting on each mound side. Also
 found is that the dependence of the mound slope on growth condition
 is due to the variation of the relative population of the facets. \par

 \section{Simulation Schemes}

    Kinetic Monte Carlo (KMC) simulation is utilized to study the thin film
 growth by deposition. In most of the previous KMC simulations,
 the deposition process is treated by randomly or uniformly
 positioning atoms at arbitrary adsorption sites. In the present
 study, when the deposition process is selected during usual KMC
 routine, an MD routine is called to simulate the trajectory of a
 deposited atom by fully considering the interaction between the
 deposited atom and substrate atoms.\par

   The details of the simulation are as follows. The substrate, Cu(001),
 lies on the xy-plane (at z=0) with x-axis lying parallel to the
 [110] direction. In the deposition process, the deposition starts at
 the height of 11-28 $a_z$ above the substrate, and deposited atoms
 are incident at an angle in the direction of x-axis. Here, $a_z$ is
 the interlayer spacing of Cu(001). The interaction between a
 deposited atom and substrate atoms is calculated by summing pairwise
 Lennard-Jones potentials, $U(r)=4D[(\sigma/r)^{12}-(\sigma/r)^{6}]$.
 Here, $D=$ 0.4093 eV, $\sigma =$ 2.338 \AA,\cite{Sanders1} and $r$
 is the distance between two atoms. The initial kinetic energy of the
 deposited atom is set to 0.15 eV, which corresponds to the melting
 temperature of copper. During each deposition process, all the
 substrate atoms are assumed to be frozen in their positions. Verlet
 algorithm is used in the MD simulation.\par

  In between two sequential deposition processes, KMC simulation is
 performed to simulate the diffusion processes of atoms on the
 substrate. In the KMC, only the diffusion into empty lattice sites
 is allowed and the exchange diffusion is not allowed. Also the
 overhang is not allowed during both deposition and diffusion
 processes. The simulation system is composed of 400$\times$400
 atomic lattice sites in fcc (001) surface and a vacuum region on top
 of the substrate with height of 28 atomic layers. \par

  \begin{table}
  \caption{ diffusion barriers and
  parameters adopted in  our simulation}
  \begin{ruledtabular}
  \begin{tabular}{cc}
  type of diffusion & diffusion barrier \\
  \tableline
  single adatom hopping(E1) & 0.48 eV\\
   step edge diffusion (E2) & 0.44 eV \\
   dimer lateral bond breaking (E3) & 0.46 eV \\
   re-estbilishing of a NN bond (E6) & 0.18 eV \\
   ES barrier (ES) & 0.10 eV\\
   ES barrier (kink site) & 0.05 eV\\
  \tableline
  jump frequency($\nu_{0}$) & $2.4 \times 10^{13} $ \\
  deposition rate ($F_{0}$) & 0.00416 ML/s \\
  \end{tabular}
  \end{ruledtabular}
  \end{table}

   Values of diffusion coefficients and diffusion barriers are adopted from
 those used by Furman and coworkers,\cite{Furman,Mehl} which
 reproduced the surface morphology of Cu islands on Cu(001) very
 well; the step Ehrlich-Schwoebel (ES) barrier is 0.10 eV and the
 kink ES barrier is 0.05 eV. In total, eleven kinds of diffusion
 barriers (including the ES barriers) are used in the KMC simulation
 and some of the important diffusion barriers are listed in Table I.
 Note that the barrier (E2) for the diffusion along step edge is
 0.44 eV in the present study,
 which is much larger than the generally accepted values of 0.2
 $\sim$0.3 eV, to save the computation time for the very frequent
 diffusions back and forth along steps.
 We examine the dependence of the surface morphology on E2,
 and find that  that the morphology
 does not show any noticeable dependence on E2 down to 0.34 eV,
 the lowest tested value of E2.
 In addition, the simulation with the E2 fairly reproduces the real
 growth mode. Hence, we anticipate that the high E2 value would not
 seriously limit the validity of the present simulation.\par

  The surface roughness is determined by the root-mean-square
  fluctuation of surface height around the mean height.
  The mound radius is determined as the radius (r) that makes the first
 zero of the height-height correlation function $G({\vec r})= <h({\vec
 r})h(0)> - <h(0)>^{2} $, and the mound radii along x- and y-axis are
 calculated separately. All the results presented are
 obtained from the average of 20 simulations under identical
 conditions. Unless mentioned otherwise, the distance in
 a plane is in unit of a$_{nn}$,   and that in the vertical direction is
 in unit of a$_z$. Here, the nearest neighbor distance, a$_{nn}$,
  is $a_{nn}=a/\sqrt{2}$, and the interlayer distance is $a_{z}=a/2$,
  where $a$ is the lattice constant of Cu. \par

 \section{RESULTS}

 \subsection{Roughness}

  \begin{figure}
  \includegraphics[width=0.45\textwidth]{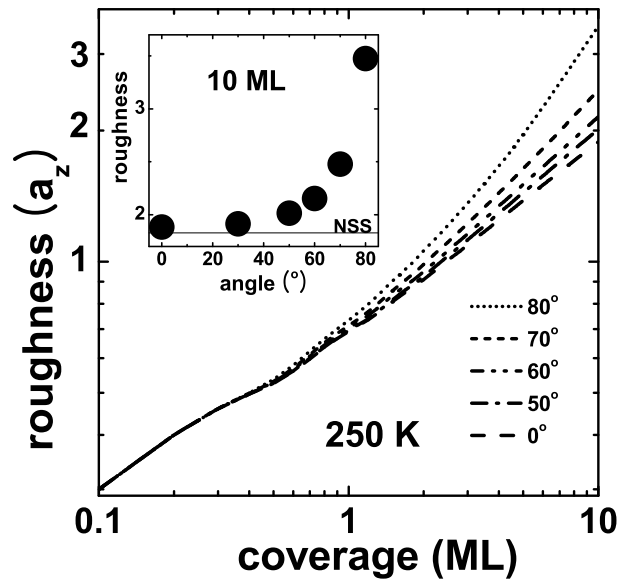}
  \caption{Surface roughness as a function of coverage at 250 K. Inset
 shows the roughness as function of the deposition angle at 10 ML
 coverage. The solid line in the inset is the simulation result of random
 deposition without considering the steering or screening effect (NSS: No
 Steering or Screening effect). }
 \end{figure}

  In Fig. 1,  the surface roughness is presented
 as a function of the coverage ($\Theta$)
  when Cu atoms are deposited at various deposition angles on Cu(001) at 250K.
  The most notable feature is that, even at the same
 coverage, the surface becomes much rougher as the deposition angle
 increases. Also, as $\Theta$ increases, the difference in roughness
 between the grazing angle deposition and the normal deposition
 ($\theta = 0^o$) multiplies. At $\Theta =$ 10 ML, the roughness with
 $\theta = 80^o$, comes to be twice larger than that with the normal
 deposition. (Inset of Fig. 1) The presently observed dependence of
 the roughness on the deposition angle is expected to originate from
 the deposition dynamics rather than the diffusion kinetics of the
 atoms on the substrate on the following grounds; (1)  such angle
 dependence of the roughness is observed irrespective of the
 substrate temperature ($T$) (Fig. 2), and (2) both steering and
 screening effects become more effective as $\theta$ increases, as
 revealed by the simulation.(Fig.1 Inset) \par

  \begin{figure}
  \includegraphics[width=0.45\textwidth]{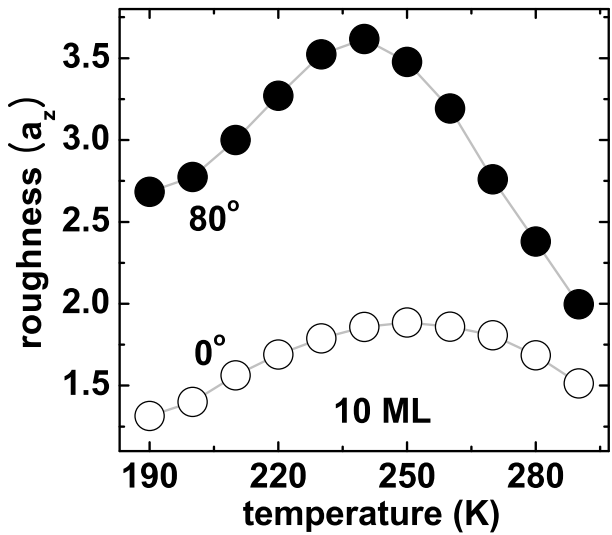}
  \caption{Surface roughness as a function of temperature at coverage of
  10 ML for two deposition angles of 80$^o$ and 0$^o$ (normal
 incidence).  The temperature dependence is more distinct for 80$^o$ case
 than 0$^o$ case. }
  \end{figure}

 In Fig. 2,  shown is the dependence of the roughness on $T$; the
 roughness tends to decrease when $T$ becomes too high or too low.
 The $bell-shape$ curve is well-explained by the $T$-dependence of
 the destabilizing current by the ES barrier.\cite{Amar} Thus, such
 $T$-dependence is caused by the kinetics of deposited atoms and is
 irrelevant to the deposition dynamics. When the deposition is made
 at $\theta = 80^o$, such $bell-shape$ $T$-dependence is also
 observed, but now in an amplified form. This illustrates that the
 surface roughness is determined $synergetically$ by the diffusion
 kinetics and the deposition dynamics.\par

 \subsection{Mound radius}

  \begin{figure}
  \includegraphics[width=0.45\textwidth]{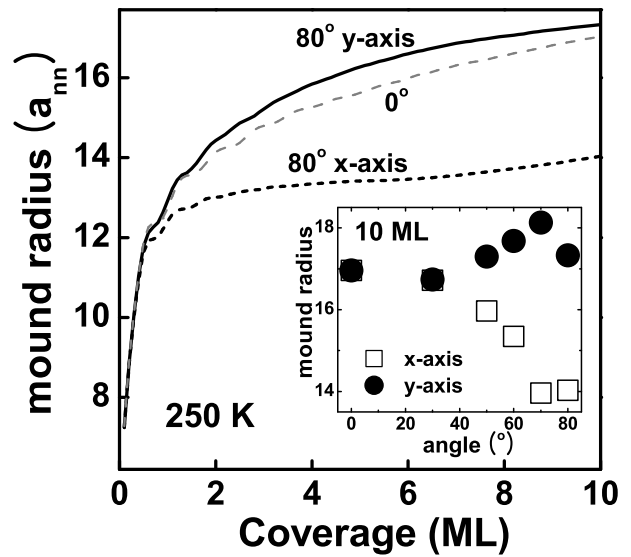}
  \caption{Mound radius as a function of coverage. Results for
 $\theta = 80^o$ are shown in solid and broken curve for
 radius along x- and y-axis, respectively. Result for the normal deposition
 ($\theta = 0^o$) is also shown in gray broken curve. The substrate
 temperature is 250 K for all the cases. Inset: the mound radius as a
 function of the deposition angle at the coverage of 10 ML.
 Open squares and
 closed circles signify the radii along x- and y-axis, respectively. }
  \end{figure}

    The mound radius as a function of the coverage and the deposition angle
 is shown in Fig. 3. When atoms are deposited in normal direction
 ($\theta = 0^o$), square mounds form with the same four-fold
 symmetry as that of the substrate. However, when the atoms are
 deposited at grazing angles, rectangular mounds form with evidently
 elongated sides along y-axis. (Here, x(y)-axis is parallel
 (perpendicular) to the deposition direction.) It is conspicuous,
 from Fig. 3 and its inset, that the difference between mound radii
 in x- and y-axis increases as does the coverage or the deposition
 angle.  This prediction agrees well with the experimental results of
 Dijken {\it et al.}\cite{Dijken2} and Lu {\it et al.}\cite{Lu},
 where  such side way growth of the mound with high aspect ratio has
 been observed with the deposition at grazing angle. In
 regards to Fig. 3, the anisotropy in the shape of mound is caused
 mainly because the growth of the mound along x-axis slows down once
 it reaches a certain length, $\simeq$13 $a_{nn}$. \par

  \begin{figure}
  \includegraphics[width=0.45\textwidth]{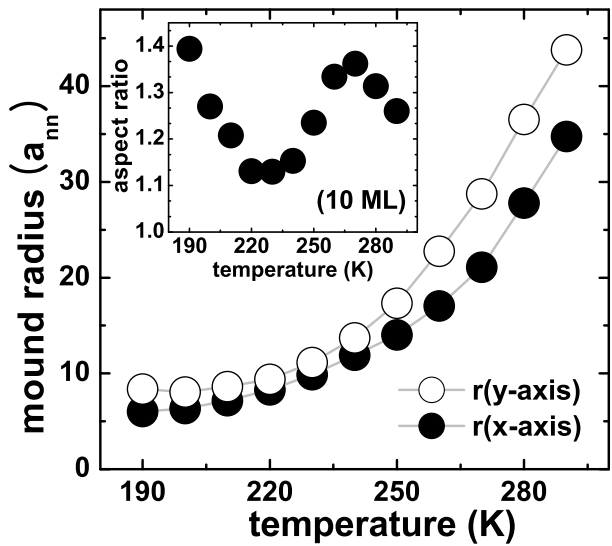}
  \caption{Mound radius as a function of temperature for the deposition angle of
 80$^o$ and the coverage of 10 ML. Open and closed circles are for
 the radius along x- and y-axis, respectively. Inset: the aspect
 ratio of the mound radii as a function of temperature.}
  \end{figure}

  Fig. 4 shows $T$-dependence of the mound radii in two
 directions when 10 ML film is grown at $\theta = 80^o$.
 It shows that both mound radii increase as does $T$
 due to the increased atomic diffusion length. At both
 high and low $T$ regimes, the mound shape is very
 asymmetric. The inset of Fig. 4 shows the aspect ratio as a function
 of $T$ at $\Theta =$ 10 ML,  and  the difference between the 
 two mound radii as high  as 40 \%. (Here, the aspect ratio is defined
 as the ratio of the mound radius along y-axis to that along  x-axis.)
  The complicated dependence of the aspect ratio on the
 temperature suggests that the diffusion kinetics also plays a
 substantial role in the determination of the mound shape in
 conjunction with the deposition dynamics. The asymmetric mound shape
  is, however, found over the whole $T$-range, which illustrates
 that the effect of deposition dynamics on the mound shape is never
 wiped out by the diffusion kinetics over the examined $T$-range.  \par

 \subsection{Mound slope}

   For the characterization of the slopes of the mound, we
 investigate the local slope
 at each step on each side of the mound that is defined as
 the step height divided by the width of the adjacent
 lower terrace. Most steps in the sides of the mounds are of
 one-atomic-layer height. Thus, if the width of a lower terrace
 adjacent to a step is 0.5 a$_{nn}$, then, the local slope is
 a$_z$/0.5 a$_{nn}$ which corresponds to the slope of \{1,1,1\}-facet
 on the fcc (001) surface. For the steps with the terrace widths, 1.5
  and 2.5 $a_{nn}$, the corresponding local slopes are those
 of \{1,1,3\}- and \{1,1,5\}-facets, respectively.

  \begin{figure} 
   \includegraphics[width=0.45\textwidth]{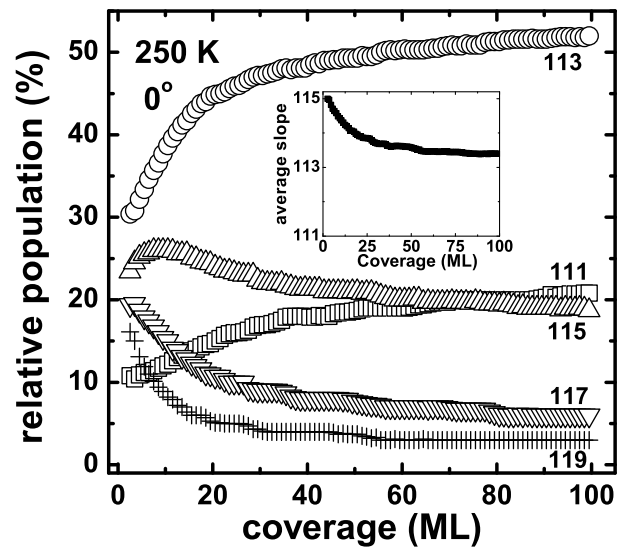}
  \caption{Distribution of the facets that compose the sides of the
 mounds formed with the normal deposition (0$^o$) at  250 K. Inset:
 the average slope of the mounds as a function of the coverage. }
  \end{figure}

 In our simulation, we find that various kinds
 of steps coexist on each side of mound.
 In Fig. 5,  the distribution of the various steps with different
 local slopes is presented as a function of the coverage for the thin
 films grown by the deposition at $\theta = 0^o$ and $T =$ 250 K. As
 $\Theta$ increases, so does the mean slope. (Inset of Fig. 5) At
 $\Theta =$ 100 ML, the mean terrace width becomes 1.67 $a_{nn}$ that
 is close to that of the \{1,1,3\}-facet. However, at the coverage,
 the relative population of the step with the local slope of
 \{1,1,j\}-facet (from now on, referred as \{1,1,j\}-step) is 21 \%
 (j=1), 53 \% (j=3), 18 \% (j=5), 6 \% (j=7), and 2 \% (j=9).
 (The portion of steps with  their slopes less than that of \{1,1,11\}-facet
 is negligible.)   Therefore, even though the mean slope converges to that of
 \{1,1,3\}-facet, the actual relative population of steps with the
 mean slope is only 50 \% and the rest of the mounds are composed of
 the steps with relatively wide range of the local slopes.  \par

  \begin{figure}
  \includegraphics[width=0.45\textwidth]{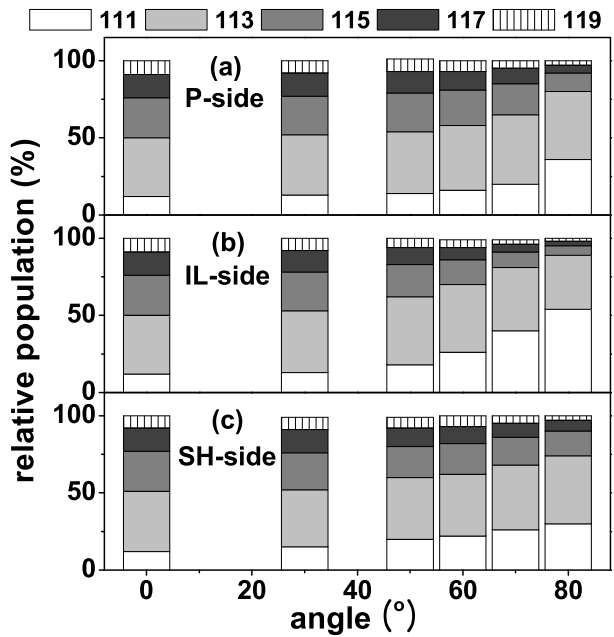}
  \caption{Distribution of the facets that compose the sides of the
 mounds formed after depositing 10 ML at 250 K at various deposition angles.
 (a) P-side: Two sides of mound facing in the direction perpendicular
 to the deposition direction, (b) IL-side: the side facing the deposition direction,
 and (c) SH-side: the shadowed (or back) side of  mound in the deposition direction.}
  \end{figure}

  \begin{figure}
  \includegraphics[width=0.45\textwidth]{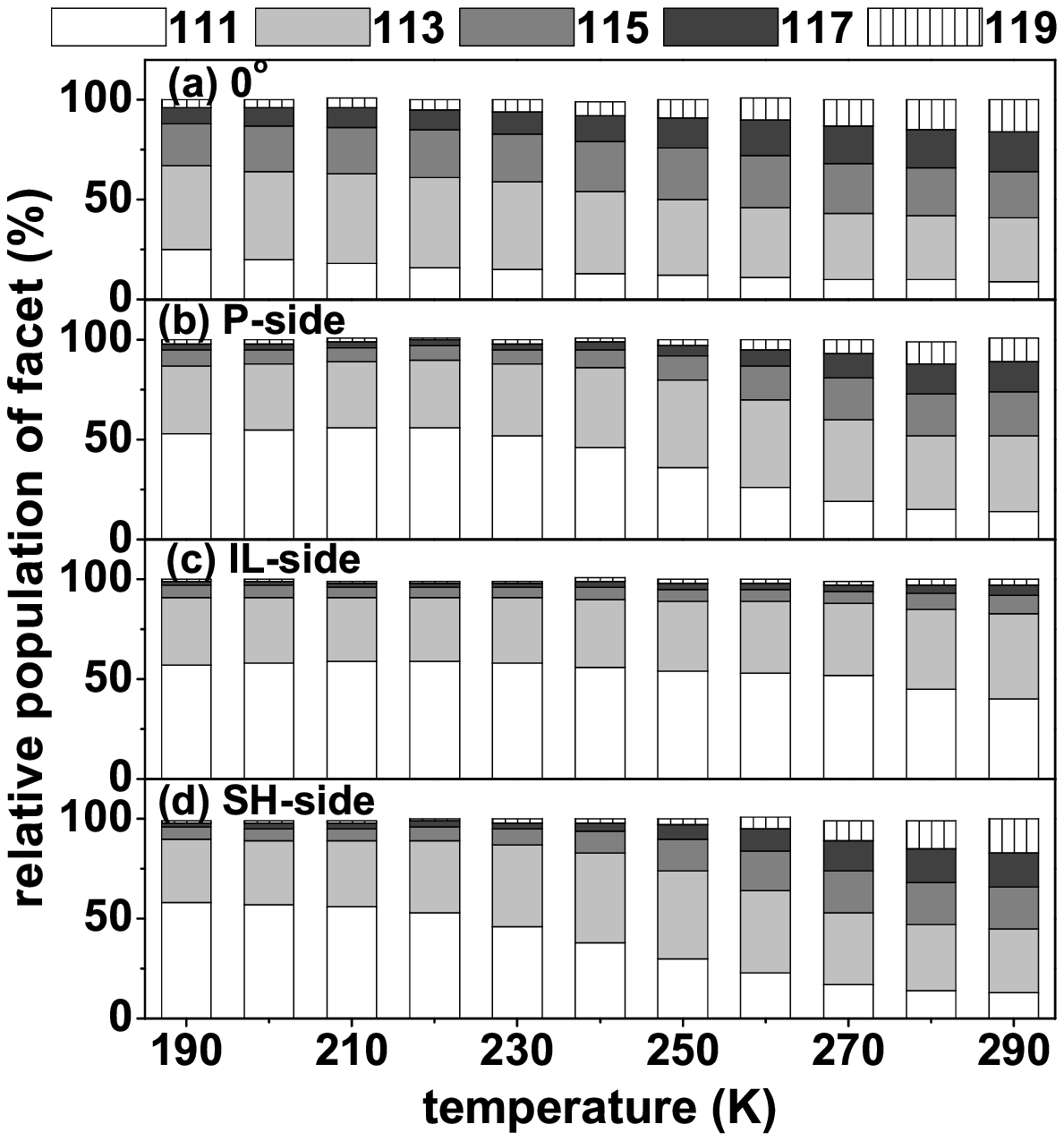}
  \includegraphics[width=0.45\textwidth]{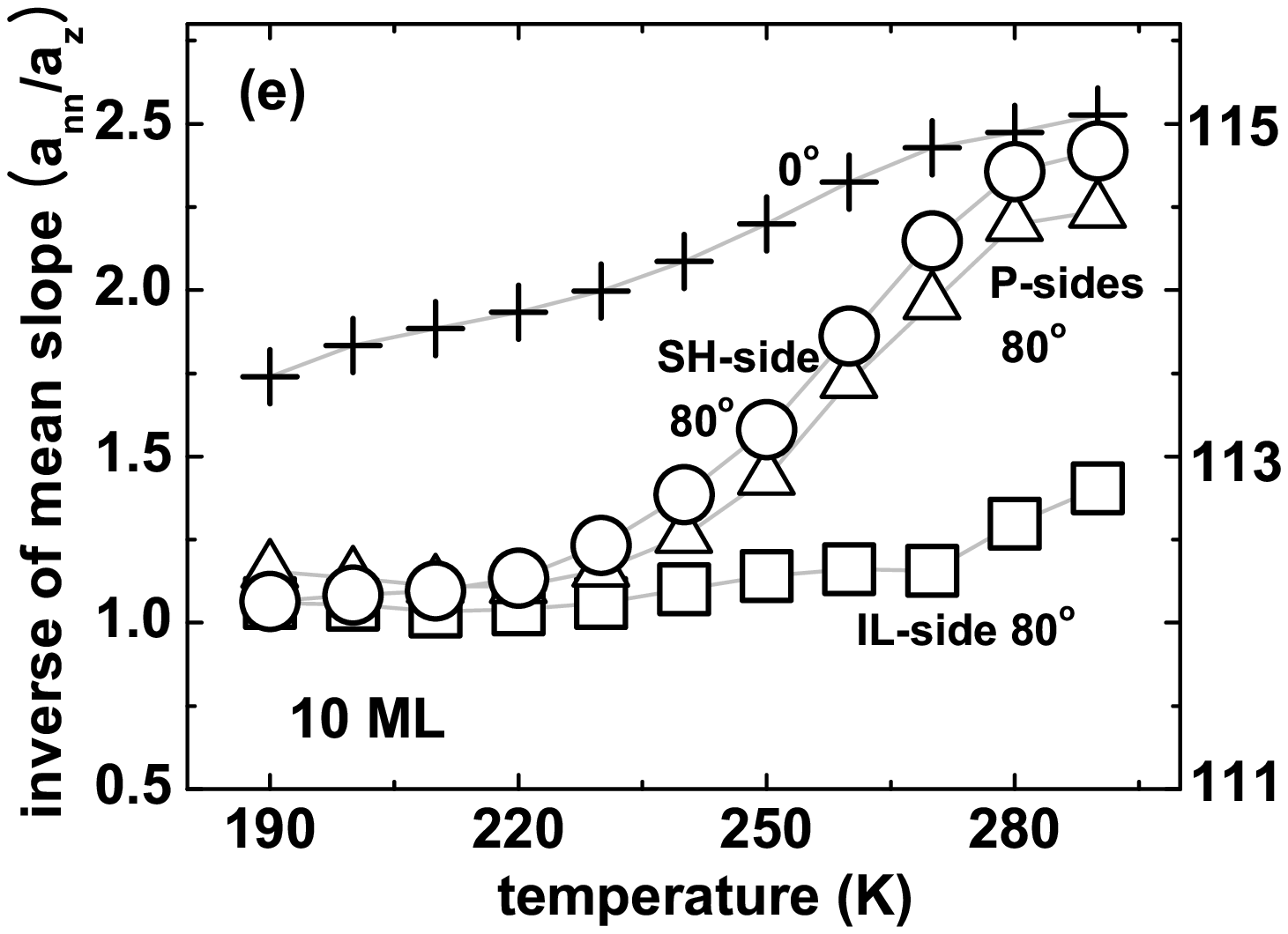}
  \caption{Relative population of various facets and average facet slope
 at the coverage of 10 ML  as
 a function of temperature for both deposition angles of (a) 0$^o$  and (b-d) 80$^o$.
 (e)Mean slopes of the three sides of the mound
 as a function of the deposition temperature for the coverage of 10 ML. }
  \end{figure}

 Now, we examine the dependence of the distribution of slopes or
 steps on the deposition angle for three inequivalent sides of the
 mounds, illuminated (IL), shadowed (SH), and perpendicular (P)
 sides. (See the caption of Fig. 6 for the description of the sides.)
 In Fig. 6 shown is the distribution of local steps on each side
 after depositing 10 ML at 250 K at various deposition angles. From
 the figure, we find that the distributions of the steps are similar
 for these three sides of the mounds for $\theta < 50^o$. As the
 deposition angle further increases, the distributions start to
 change and become quite distinct from each other at $\theta = 80^o$.
 For $\Theta =$ 10 ML and $\theta = 80^o$, 54 \% of the IL-sides are
 composed of \{1,1,1\}-step. On the other hand, only 30 \% of SH-side
 is composed of the \{1,1,1\}-step,  and thus its mean slope is much
 less steep  than that of the IL-side. In the SPA-LEED experiment of
 Cu/Cu(001), Dijken {\it et al.} \cite{Dijken2} reported that the
 IL-side is made of \{1,1,1\}-step and SH- and P-sides are of
 \{1,1,3\}-step. These experimental results nearly match the
 average behaviors observed in our simulation. \par

   We also investigate the distribution of the steps
   as a function of the substrate temperature (Fig. 7).
   For the case of grazing angle
 deposition, the step distributions (Fig. 7 (b)-(d)) and the mean
 slopes(Fig. 7 (e)) on the three sides become similar around 220 K.
 Further, they vary little, as the substrate temperature is lowered
 below 220K. These observations suggest that the step distribution
 for $T < 220 K$ represents the limiting behavior of the slope
 formation or the steepest slope reachable by deposition, during
 which some less steep steps such as \{113\}-steps other than the
 steepest \{111\}-steps occurs due to statistical fluctuation.\par

 As $T$ increases, the portion of the less steep steps increases and
 that of the steeper steps such as \{111\}-step decreases. That is,
 the mound becomes smooth as $T$ increases. The way of smoothing,
 however, differs depending on the mound side: As $T$ increases, the
 P-side (Fig. 7(b)) and the SH-side (Fig. 7(d)) rapidly become
 smooth, while the change in the step distribution on the IL-side
 occurs in the smaller scale (Fig.
 7(c)). This can also be seen clearly in the $T$ dependence of the
 mean slopes in Fig. 7(e). \par

 \section{DISCUSSION}

 \subsection{Effects of the deposition dynamics on the morphology of the film}

 In the previous section, we observe that the roughness (Inset of
 Fig. 1 and 2), the mound shape (Fig. 3 and 4) and the mound slope
 (Fig. 5-7) depend on the non-kinetic variable such as deposition
 angle. Such angular dependence can be attributed to the deposition
 dynamics that includes (1) the steering of the trajectory of the
 deposited atom due to  its interaction  with substrate
 atoms and (2) the screening effects of the deposited atoms due to
 the geometrical structure already formed on the substrate. Both
 effects cause inhomogeneous deposition flux
 depending  on the deposition condition and make the growth of thin films
 sensitive to the deposition condition. \par

  \begin{figure}
  \includegraphics[width=0.45\textwidth]{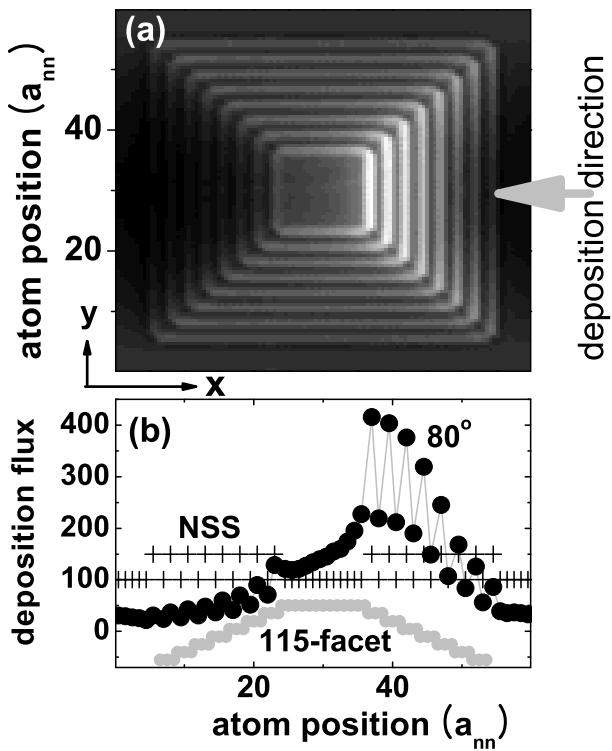}
  \caption{Deposition flux calculated by MD simulation.
 Atoms are deposited at grazing deposition angle of 80$^o$ on an 8-layer
 high mound surrounded by (1,1,5)-facets. (a) The local deposition flux in
 gray scale, where the brighter color indicates the higher local flux. (b) The local
 deposition flux along the line crossing the center of the mound along x-axis.
 The ordinate is the deposition flux in percentage
 relative to the average deposition flux over the total area. The deposition
 flux at normal (0$^o$) deposition is shown with + symbol as a
 reference and the mound is depicted by gray circles at the bottom. }
 \end{figure}

 To study the effects of the inhomogeneous deposition flux on the
 growth of thin film, we investigate the flux on an 8-layer high
 mound surrounded with \{1,1,5\}-facets by MD simulation. Fig. 8(a)
 shows the deposition flux over the mound in gray scale for $\theta =
 80^o$. Fig. 8(b) shows the deposition flux along a line passing
 through the center of the mound along the x-axis, which shows strong
 asymmetry, i.e., the deposition flux on the IL-side is 2 to 4 times
 larger than the average deposition flux and that on SH-side is only
 about 10 to 50 \% of the average. The simulation with the mounds
 surrounded by the different facets show the same trend, $i.e.$, the
 enhanced deposition flux on the IL-side and the reduced deposition
 flux on the SH-side.\par

 Such inhomogeneous deposition flux gives rise to different growth
 speeds on each side, and thus the film is rougher than the film
 grown under the normal deposition condition (Figs.1 and 2).
 Especially, the enhanced deposition flux near the front edge of the
 top terrace increases the destabilizing current toward ascending
 step edge by the ES barrier, and critically roughens the
 surface.\cite{Amar}(See the following section for the detailed
 description.) \par

 The  asymmetric shape of mound formed during grazing angle
 deposition (Fig. 3 and 4) is attributed to the decrease of the
 overall deposition flux along x-direction (i.e., over both the IL-
 and SH-sides), and the simultaneous increase of the effective
 deposition flux over the P-sides (Fig. 8) due to the following two
 reasons; the first one is the net mass transfer from the region near
 the rear edge of the mound to the top of the mound due to the
 steering effects, reducing the deposition flux along
 x-axis.\cite{Dijken1,Seo1} Some of the transferred mass on the top
 of the mound is in the long run redistributed equally to the four
 sides.\cite{Seo1} It effectively increases the deposition flux along
 the y-axis and {\it vice versa}. The second one is the increased
 deposition flux over the P-sides due to the attraction of the
 deposited atoms moving along the edges of the P-sides toward the
 sides.\cite{Seo1} Such imbalance of the effective deposition flux
 results in the faster growth speed in the y-direction than in the
 x-direction, giving birth to the asymmetric mounds elongated 
 along y-axis. \par

 The observed asymmetric slopes of the mounds observed for $\theta >
 50^o$ (Fig. 6) can also be explained by the inhomogeneous deposition
 flux on the following two grounds. Firstly, the higher deposition
 flux on IL-side than that on SH-side offers growth
 environment effectively equivalent to the lower growth temperature
 on IL-side than that on SH-side. Thus, the terrace size in
 IL-side is narrower than that in SH-side or, equivalently, the slope
 in IL-side is steeper than that in SH-side as observed in both the
 previous experiment\cite{Dijken2} and the present simulation. \par

 In addition, the flux distribution on the top terrace strengthens
 the asymmetric slope formation; A notable feature of the deposition
 flux in Fig. 8(b) is that the flux near the edge toward IL-side is
 much higher than that near the opposite edge, which should result in
 the increased density of islands near to the edge of IL-side
 on the top terrace. Sequential formation of islands on the top terrace
 preferentially close to the edge makes the mound have steps
 with the narrower terrace width or the steeper slope on IL-side than
 those on SH-side. \par

 In Fig.7, we observe the dependence of the smoothing kinetics on the
 side of the mound; as $T$ increases, relatively rapid smoothing
 occurs on both the P- and SH- sides, while it is retarded at the
 IL-side. This is because the deposition flux is larger on the
 IL-side than those on the other sides (Fig. 8); The mean capture
 length of the deposited atoms to form islands on the IL-side is
 shorter than those on the other sides. It means that the effective
 temperature felt by the deposited atoms on the IL-side is lower than
 that on the other sides. As a result, the smoothing proceeds
 relatively slowly on the IL-side, as $T$ increases.  \par

 In summary, both (1) the inhomogeneous deposition flux over the
 sides of the mound due to the steering and screening effects and (2)
 the enhanced deposition flux near the front edge of the top terrace
 of the mound due to the steering effect cooperatively give rise to
 the asymmetric mound formation with different lengths and slopes on
 each side,  and also accelerate the roughening of surface during
 deposition at grazing angle.\par

 \subsection{The steering effects v.s. the screening effects}

  \begin{figure}
 \includegraphics[width=0.45\textwidth]{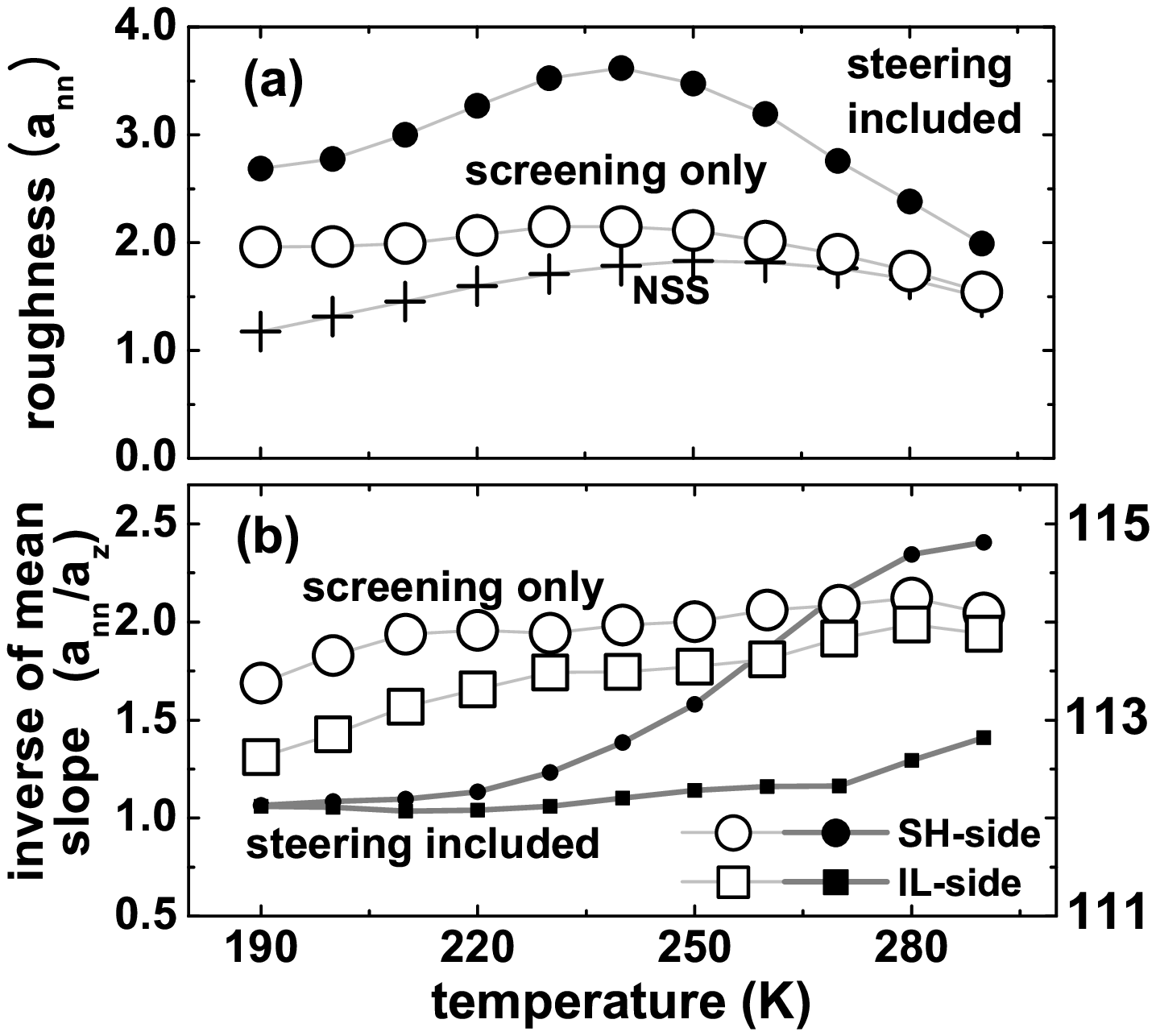}
 \includegraphics[width=0.45\textwidth]{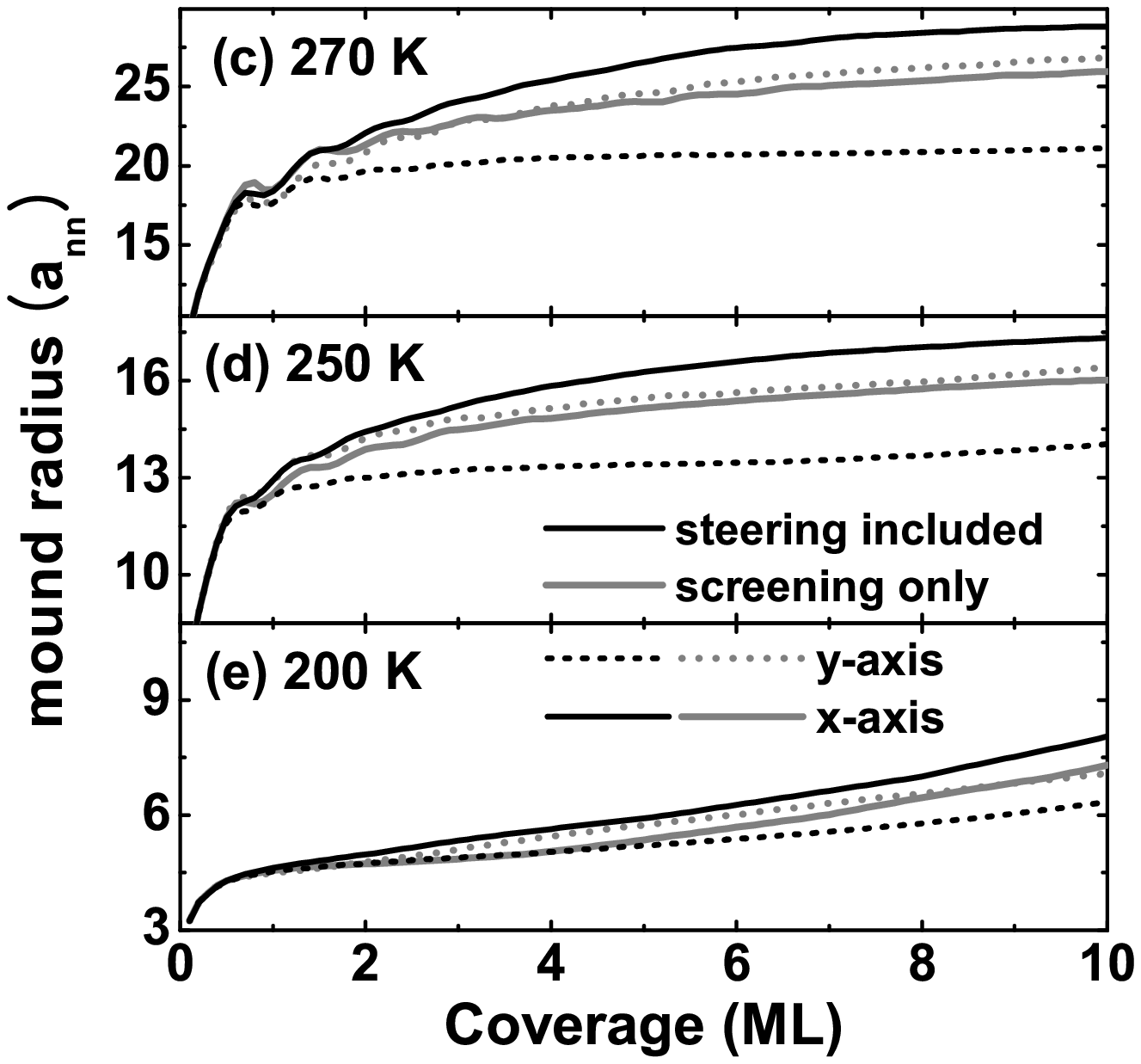}
   \caption{Simulation results with the screening effects alone
 with $\theta = 80^o$ and $\Theta = 10$ ML. As a  reference, the
 simulation results considering both screening and steering effects
 are also presented. (a) Roughness as a function of temperature: open
 circles for deposition considering screening effects alone, + for
 random deposition (NSS), and closed circles for deposition
 considering both screening and steering effects. (b) Mean slopes for
 the SH-side (circle) and the IL-side (square). Open (closed) symbols
 for deposition including screening effects alone (both screening and
 steering effects). (c-e) Mound radius as a function of coverage.
 Gray (black) curves correspond to the deposition considering
 screening effects (both screening and steering effects). Solid and
 broken curves are radii along x- and y-axis, respectively. }
 \end{figure}

  As the thickness of the film become thick, in addition to the
 steering effects, the geometric screening effects play an important
 role. To study the contribution of the individual dynamic effect to
 thin film growth, we carry out another set of growth simulation
 which takes only the geometric screening effects into account. In
 this simulation, the trajectory of the deposited atom is a straight
 line determined by the initial position and velocity of the
 atom.\par

 In Fig. 9(a) shown is the roughness evolution as a function of the
 temperature for three different cases; (1) the case assuming no
 steering and screening effects (NSS) or random deposition, (2) the
 case taking only the geometric screening effects into account, and
 (3) the usual simulation taking all the dynamic effects, i.e., both
 steering and screening effects. At low temperature (e.g., 190 K),
 the screening effects contribute to the roughening almost as much as
 the steering effects. Here, the contribution of the screening
 effects to the roughening is estimated as the difference between the
 roughness obtained with the screening effects included and that of
 NSS. The contribution of the steering effects is estimated similarly
 as the difference between the roughness with both effects considered
 and that with only the screening effects. The roughening due to the
 screening effects, however, decreases gradually as temperature
 increases and becomes negligible at temperatures higher than 260 K.
 On the other hand, the roughness due to the steering effect  keeps
 increasing up to 240 K and starts to decrease gradually at the
 higher temperatures. This suggests that the effects of the
 inhomogeneous deposition flux due to the screening effect alone
 relaxes through diffusion kinetics at a temperature lower than that
 due to the steering effects.\par

  \begin{figure}
  \includegraphics[width=0.45\textwidth]{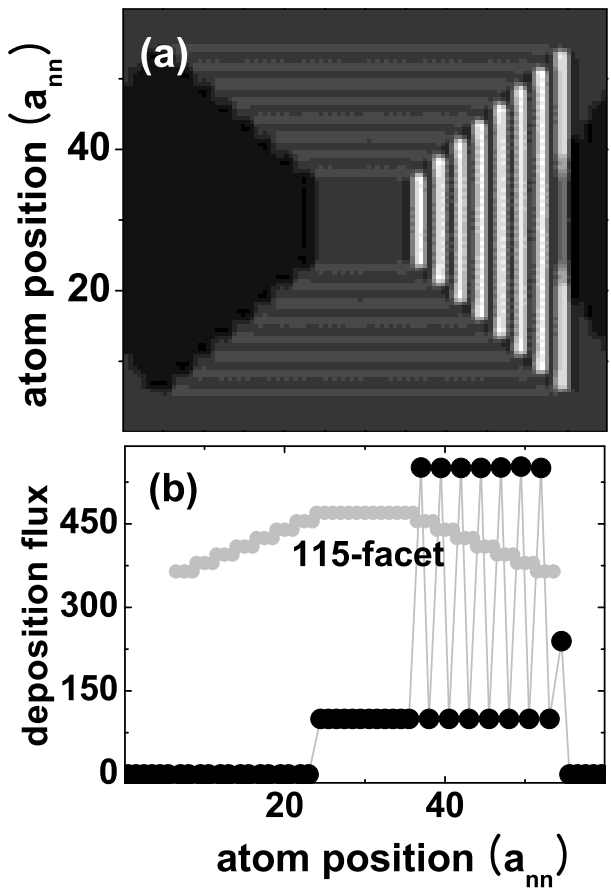}
  \caption{Deposition flux calculated by considering only the screening
 effects. Atoms are deposited at grazing deposition angle of 80$^o$
 on an 8-layer high mound surrounded by (1,1,5)-facets.
 (a) The local deposition flux in
 gray scale, where the brighter color indicates the higher local flux.
 (b) The local
 deposition flux on the line crossing the center of the mound along
 x-axis is presented; the ordinate is the relative amount (\%) of
 deposition flux relative to the average deposition flux over the
  whole surface. The mound is shown with gray circles in the upper region.}
 \end{figure}

 The origin of the difference in the roughness and the relaxation
 temperature between these   two cases lies in their different
 flux distributions: In Fig. 10 shown is the flux distribution for the same mound as
 that in Fig. 8, but only with the screening effects taken into
 account. The flux near SH-side is totally depleted and about the same
 amount is added to IL-side. The most notable difference between
  two cases is found in the flux on the top terrace. The flux distribution
 taking both dynamic effects in Fig. 8 shows pronounced enhancement
 of the flux near the front edge of the top terrace, while no such enhancement of flux
 is found in Fig. 10. The steering effect causes
 vertical mass redistribution, displacing the flux
 on SH-side expected  for the random deposition to the top terrace,
 accelerating the roughening of the surface. In contrast, the
 screening effect just redistributes
 the depleted flux from SH-side to IL-side in the same plane.
 Thus, for the relaxation of the screening effects,
 only the diffusions in the same plane via terrace
 diffusion or diffusion along the edges of mound
 need to be activated.   For the relaxation of the steering effects, however,
  interlayer diffusion across the ES barrier that requires much higher activation energy
 than that for the in-plane diffusion should be accompanied. In fact,
 the $bell-shape$ dependence of the roughness on the temperature in
 Fig. 9(a), shown for the case when the steering effect is also
 considered, is reminiscent of the growth mode where the limiting
 process is the diffusion over ES barrier.\cite{Amar} Hence, the
 relaxation of the steering effects takes place at temperature
 higher than that for the relaxation of the screening effects. \par

   In Fig. 9(b), the mean slopes for IL- and SH-sides are shown as a function
 of growth temperature also for the aforementioned two cases. The
 mean slope of the IL- side is steeper than that of the SH-side over
 the whole temperature range for both cases. If only the screening
 effect is considered, the difference in their slopes dwindles as $T$
 increases, and finally becomes almost negligible around 270 K. At
 250 K, the slopes of both sides are slightly lower than that of
 \{113\} facet. In the previous experiment depositing 40 ML of Cu on
 Cu(001) at 250 K\cite{Dijken2}, the slopes of the IL- and SH-side is
 found to be those of \{111\} and \{113\}, respectively, in
 contradiction with the results of the aforementioned simulation
 considering the screening effects alone.\par

 The inclusion of the steering effect makes the slopes steeper,
 especially  on the IL-side, and the difference in
 slopes becomes evident as observed in the experiment. (Fig. 9(b))
 Further,  the mean slope of the IL-side lies in between \{111\} and
 \{113\}, and that of the SH- side corresponds to that of \{113\}.
 These slopes now approach the ones observed in the previous
 experiment.\cite{Dijken2}  \par

 Regarding the mound radius, the screening effects alone make
 the mounds of almost symmetric shape (Figs. 9 (c)-(e))
 which strikingly differs
 from what has been observed in the experiments.\cite{Dijken2,Lu}
 This is understood by the fact that the flux
 blocked by the screening effect on the SH-side
 results in the increase of the flux on the IL-side (Fig. 10),
 which looks as if  the blocked flux is simply displaced
 to the IL side.  Therefore,  the reduced lateral growth speed on the SH-side is
 almost compensated by the increased growth speed
 on the IL-side.
 Hence, the overall shape of mound remains square symmetric,\cite{square} if only the
 screening effects are taken into account in the growth.
  The asymmetric mounds with
 longer side along y-axis (Fig. 9)  form as observed in the
 experiment,\cite{Dijken2} only when the steering effect is
 added.  \par

   In short, in the thin film growth at grazing deposition angle, the
 asymmetric mound shape, the surface roughness, and the slope
 difference on the IL- and SH-sides of mounds
 result mainly due to
 the steering effect rather than the screening effect. (Fig. 9).
 Remembering that the main difference of the steering effects from
 the screening effects is the higher deposition flux near the front
 edge of the top terrace, the observed characteristics of the films
 grown at grazing deposition angle
  is largely determined by the growth characteristics of the
 top layer as the growth front.  \par

 \section{SUMMARY AND CONCLUSION}

  We perform KMC simulation to study the thin film growth by deposition at
 grazing angle. We observe (1) the notable increase of the surface
 roughness and (2) the asymmetry in both mound shape and slopes, as
 compared with the thin film grown by the normal deposition. Such
 results are in good agreement with the previous experimental
 observations. We find that the aforementioned structural features of
 the films grown by grazing angle deposition are mainly attributed to
 the steering effect rather than the screening effect. Especially,
 the inhomogeneous deposition flux on the top terrace induced by the
 steering effects is the most influential factor. \par

 We also make an additional interesting observation that the mound
 side is not composed of one kind of facet, even when the slope selection is attained.
  Instead, we find that there coexist
 variety of local facets and the selected mound slope observed in
 experiment represents only the mean slope of those.
 Therefore, the slope selection does not mean the facet selection.  \par

 \begin{acknowledgements}

 \end{acknowledgements}


\begin{references}
\bibitem{square} The slightly longer mound length in the y-axis
 than in the x-axis may be
 caused by the net mass transfer through diffusion from the IL-side
 to the P-side due to the large adatom density gradient resulting from
 the large difference in the deposition flux between the two sides.
\end{references}
  \end{document}